\documentclass[aps,prl,showpacs,twocolumn,amsmath,amssymb,superscriptaddress,footinbib]{revtex4}

\usepackage[english]{babel}
\usepackage{latexsym}
\usepackage{graphics}
\usepackage{subfigure}
\usepackage{epsfig}
\usepackage{color}

\def\be{\begin{equation}}
\def\ee{\end{equation}}
\def\bea{\begin{eqnarray}}
\def\eea{\end{eqnarray}}
\def\bi{\begin{itemize}}
\def\ei{\end{itemize}}
\def\bin{\begin{enumerate}}
\def\ein{\end{enumerate}}

\newcommand{\ket}[1]{| #1 \rangle}

\begin{document}

\title{Disorder-Induced Order in Quantum $XY$ Chains}

\author{A.~Niederberger} \affiliation{ICFO-Institut de Ciencies Fotoniques, Mediterranean Technology Park, 08860 Castelldefels (Barcelona), Spain}
\author{M.~M.~Rams} \affiliation{ICFO-Institut de Ciencies Fotoniques, Mediterranean Technology Park, 08860 Castelldefels (Barcelona), Spain} \affiliation{Marian Smoluchowski Institute of Physics and Mark Kac Complex Systems Research Centre, Jagiellonian University, Reymonta 4, PL-30059 Krak\'ow, Poland} \affiliation{Theoretical Division, Los Alamos National Laboratory, MS-B213, Los Alamos,NM 87545, USA}
\author{J.~Dziarmaga} \affiliation{ICFO-Institut de Ciencies Fotoniques, Mediterranean Technology Park, 08860 Castelldefels (Barcelona), Spain} \affiliation{Marian Smoluchowski Institute of Physics and Mark Kac Complex Systems Research Centre, Jagiellonian University, Reymonta 4, PL-30059 Krak\'ow, Poland}
\author{F.~M.~Cucchietti} \affiliation{ICFO-Institut de Ciencies Fotoniques, Mediterranean Technology Park, 08860 Castelldefels (Barcelona), Spain}
\author{J.~Wehr} \affiliation{ICFO-Institut de Ciencies Fotoniques, Mediterranean Technology Park, 08860 Castelldefels (Barcelona), Spain} \affiliation{Department of Mathematics, The University of Arizona, Tucson, AZ 85721-0089, USA}
\author{M.~Lewenstein} \affiliation{ICFO-Institut de Ciencies Fotoniques, Mediterranean Technology Park, 08860 Castelldefels (Barcelona), Spain} \affiliation{ICREA - Instituci\`o Catalana de Ricerca i Estudis Avan{\c c}ats, E-08010 Barcelona, Spain}

\date{\today}

\begin{abstract}
We observe signatures of disorder-induced order in 1D $XY$ spin chains with an external, site-dependent uni-axial random field within the $XY$ plane. We numerically investigate signatures  of a quantum phase transition at $T=0$, in particular an upsurge of the  magnetization in the direction orthogonal to the external magnetic field, and the scaling of the block-entropy with the amplitude of this field. Also, we discuss possible realizations of this effect in ultra-cold atom experiments.
\end{abstract}


\maketitle
Disorder is common in a wide range of physical phenomena \cite{books}. 
It typically originates from inhomogeneous environmental conditions and other parameters beyond experimental control. The time scale $\tau$ of observations
on the considered system makes it possible to divide disorder into two categories:
 {\it Annealed disorder}, that changes in time scales much  shorter than $\tau$, 
 and correspond to fluctuations (thermal, quantum or externally driven),
 and {\it quenched disorder}, i.e. disorder that is permanent or does not change significantly on 
the scale of $\tau$.
In general, disorder tends to reduce material properties that depend on regularity, 
such as magnetization, mechanical strength, or electric conductivity. 
However, even if counterintuitive, sometimes the opposite is true: 
there exist systems that develop order (as measured by some order parameter) when in presence
of disorder. Various instances of such counterintuitive effect occur in the context of annealed disorder 
in geometrically frustrated systems, such as frustrated antiferromagnets. The term {\it order-by-disorder} 
was introduced by Villain {\it et al.} \cite{villain} to describe the effect that occurs in the Ising model on a 2D 
square lattice with ferromagnetic interactions between all next neighbors, except for antiferromagnetic interactions 
along the, say,  even columns. Classically, such a system is truly disordered at temperature $T=0$, 
and exhibits a large number of degenerate ground states. Thermal fluctuations at arbitrarily low $T$ induce 
ferrimagnetic order, in which spins at odd columns order parallelly. 
The same ordering occurs in the case of quenched disorder induced by site dilution 
(i.e. introduction of non-magnetic sites). 
The phenomenon of order-by-disorder persists also in quantum systems, 
although its appearance is not as dramatic (for a review see \cite{misguich}).

A completely different mechanism, called {\em disorder-induced order} (DIO), has been recently investigated  by some of 
us \cite{Wehr2006}, and earlier by several other authors \cite{olderpapers}. This phenomenon 
depends crucially on two ingredients: the unperturbed system must have a 
continuous symmetry, and the disorder must break it --- for instance by promoting a single 
direction in space. Let us explain this through an example. 
Consider a chain of spins with isotropic nearest neighbor interaction in the $XY$ plane, 
which has a $U(1)$ rotational symmetry about the $Z$ axis. 
As it is well known,  the Mermin-Wagner-Hohenberg (MWH) theorem \cite{mwh} states that 
there is no spontaneous magnetization at any positive temperature for the bare (i.e. without quenched disorder) 
$XY$ model for dimensions two and smaller (in the limit of infinite systems). 
The continuous symmetry of the model is pivotal for the proof of the MWH theorem, since it is responsible for the freedom 
of fluctuations, or, more precisely, for the density of states of low energy excitations in the system. 
Let us add now to the system a quenched disorder in the form of local  external magnetic fields
 in the $XY$ plane, with random strengths and directions. 
The key ingredient for the appearance of DIO is the relation between the disorder 
probability distribution and the symmetries of the physical system.
In particular, if  a disorder distribution has  the same symmetry as the system, the disorder suppresses
the spontaneous magnetization even more: 
even arbitrarily small disorder of this form  destroys the order at $T=0$ in 2D,  
and it presumably lowers the critical temperature of the transition in 3D \cite{Aizenman1989f}. 
However, if the random field is aligned
only along the $X$ axis, the disorder
has a symmetry different than that of the system -- the MWH theorem no longer applies, 
and spontaneous magnetization can appear precisely because of the presence of disorder. 

The  mechanism of DIO is quite general -- it can be applied
whenever there is a continuous symmetry in the system that can 
be broken with a random field with a lesser symmetry. 
Indeed, we have recently showed that a disordered random field with a 
symmetry different than that of the system may increase order in the direction where there is no disorder
\cite{Wehr2006}.
We considered a two-dimensional array of classical spins with a ferromagnetic interaction
in presence of a random external magnetic field in the $Y$ direction. 
With this choice of disorder, the system has a remaining mirror symmetry 
along the $Y$ axis (which can easily be broken by appropriate boundary conditions). 
We argued that, in this case, the disorder will always induce a non-zero magnetization 
in the $X$ axis -- i.e. in the direction orthogonal to the random field. 
Similarly, the relative phase of two-component  interacting
Bose-Einstein Condensates can be fixed by applying a suitable random coupling Raman field
\cite{Niederberger2008}. 
The same can be done with the relative phase between the superfluid order parameter
of an ultra-cold Fermi gas and the wave function of a molecular BEC of the same species
\cite{Niederberger2009}. In this case the disorder is provided by a 
spatially random photo-associative-dissociative coupling between the
BEC and the Fermi gas atoms.

In this paper we report numerical studies of one-dimensional quantum spin chains that present 
disorder-induced order. In particular, we consider isotropic $XY$ spin chains subject to a random
magnetic field along one axis on the $XY$ plane. We show evidence of the formation of
spontaneous magnetization, and a transition from that phase to a phase without long-range order for a critical strength of the disordered field. Apart from the magnetization, we study the behavior of entanglement.
Using numerical methods, we are able to identify the quantum critical point for a particular case of staggered magnetic field and the critical exponents in that situation.
The transition that we discuss here is  an analogue of the one that   
can be obtained with staggered magnetic fields 
in the gapless phases of the $XXZ$ model \cite{Neresayan1994}.  
The main difference is that we mostly consider magnetic fields that oscillate on a 
much larger spatial  scale, or are fully random.  

We envision an implementation of our ideas
using ultra-cold atom experiments. 
Technological developments have made 
ultra-cold gases a powerful tool to study condensed matter systems
with an unprecedented degree of control -- allowing the manipulation of 
geometry, dimensionality, and even the
interaction strength between particles (by means of Feshbach resonances) \ \cite{Lewenstein2007,Bloch2008,Jaksch-toolbox}.
An example of the power of ultra-cold atom systems as a test bed for condensed matter
physics was the recent observation of matter-wave Anderson Localization
\cite{Roati2008,Billy2008} -- predicted already in 1958
in the context of the metal-insulator transition \cite{Anderson1958}, and for ultra-cold atoms in 2003 \cite{damski}.
These experiments, and others, illustrated an 
interesting difference between cold gases and usual condensed matter systems:
because the potentials are exceedingly clean, 
and thermal fluctuations negligible, 
disorder in cold atom experiments does not occur naturally and has to be introduced {\em by design}.
This, in turn, translates into 
complete control over the disorder distribution: e.g. 
by changing light intensity the disorder strength can be varied continuously. Typically
one can use laser speckles to induce static unstructured randomness 
(pioneered by the late G. Grynberg \cite{Guidoni1997,Grynberg2000}), 
or  superposed lasers of incommensurate wavelengths to create 
quasi-randomness (for  recent reviews see \cite{reviews-disorder}). Moreover, one can control nonlinear interaction strengths using Feschbach resonances, and reach in this way various regimes \cite{Lugan2007} of the Anderson-Bose glass phase \cite{Fisher1989}. This manipulability is essential for 
deep exploration of disorder-induced order phenomena, and for observation of the phenomena discussed in this paper.

\section{Model description}
We consider a ferromagnetic spin chain with $N$ spins $1/2$
in a random external magnetic field, described by the following hamiltonian: 
\begin{equation} \label{eq:hamiltonian}
\hat H = - \sum_{i=1}^{N-1} \left( \hat \sigma_x^i \hat \sigma_x^{i+1} + \hat \sigma_y^i \hat \sigma_y^{i+1} \right) - \sum_{i=1}^N h^i \hat \sigma_{\vec{n}}^i,
\end{equation}
where $\hat \sigma_\alpha^i$ are the $\alpha=x,y$-Pauli spin matrices at site 
$i$, and $h^i$ is the random field at site $i$.
The field points along an arbitrary direction $\vec{n}$ inside the $XY$ plane and 
$\hat \sigma_{\vec{n}}=\vec{n}\cdot\vec{\sigma}$.
Also within the $XY$ plane, we will distinguish observables (and measurements)
aligned about the axis $\vec{n}$ with a $\parallel$ subscript,
and observables perpendicular to $\vec{n}$ with a $\perp$ subscript.
As mentioned before, the second term of Eq.~(\ref{eq:hamiltonian}) does not have the
same symmetry as the first term (which is invariant with respect to rotations along the $Z$ axis).

The relevant order parameters are the mean expectation values of the magnetization
along the parallel and orthogonal directions:
$\bar m_\parallel = \langle \sum \sigma^i_\parallel \rangle/N$ and 
$\bar m_\perp = \langle \sum \sigma^i_\perp \rangle/N$.
Typically, we also consider the local magnetization 
$m_\parallel = \langle \sigma^i_\parallel \rangle$ and 
$m_\perp = \langle \sigma^i_\perp \rangle$,
indicating which regions of the chain are being discussed.
Both $m_\parallel$ and $m_\perp$ vanish as the amplitude 
of the external fields approaches zero.
For large field intensity, $m_\parallel$ follows the local direction of the field. 
In this case, the average $m_\perp$ is essentially zero.

Entanglement is known to be a good predictor of quantum phase transitions \cite{Osterloh2002,Cardy,EisertAreaLaws}.
Although there is a variety of possibilities, the observation of a singularity in an entanglement
measure most certainly implies a second order quantum phase transition. In order to measure entanglement we will use the block entropy $S(p)$,
defined as the Von Neumann entropy of the reduced density matrix
obtained by tracing out the degrees of freedom of $N-p$ spins of the chain.
By means of a Schmidt decomposition, any pure state $| \psi \rangle $ of the system
can be expressed as 
\begin{equation} \label{eq:SchmidtDecomposition}
| \psi \rangle = \sum\limits_i \lambda_i^{1/2} | \psi_i^{[1 \dots p]} \rangle \otimes | \psi_i^{[p+1 \dots N]} \rangle,
\end{equation}
where $\{| \psi_i^{[1 \dots p]} \rangle\}$ and $\{ | \psi_i^{[p+1 \dots N]} \rangle \}$ are
orthonormal states in the Hilbert space of the first $p$ and last $N-p$ spins respectively. 
Because of the orthonormality property,
in this basis it is easy to write down the reduced density matrix for the first $p$ spins,
$\rho_p = \sum\limits_i \lambda_i | \psi_i^{[1 \dots p]} \rangle \langle \psi_i^{[1 \dots p]}|$.
 The positive numbers $\lambda_i$ are the so called Schmidt coefficients, 
 and give the block entropy 
\begin{equation} \label{eq:Sofp}
S(p) = - \sum\limits_i \lambda_i \textrm{log}_2 \lambda_i.
\end{equation}
The value of $S(p)$ depends on both classical and quantum correlations (such as entanglement) between the two blocks $[1,\dots,p]$ and $[(p+1),\dots,N]$ of the $N$-spin chain. 
The so called area law says that the block entropy of a ground state generally scales with the size of the boundary (area) of the system \cite{EisertAreaLaws} --- except at criticality, where there are typically logarithmic corrections.
In one dimensional systems, the boundary of a block is constant. Thus, away from the critical point, entropy saturates beyond a certain block size $p_0$: $S(p)=S(p_0)$ for all $p>p_0$. 

Let us analyze the behavior of the block entropy of our system for some limit cases. 
For small fields, the $XY$ term in Eq.~(\ref{eq:hamiltonian}) dominates, and the system exhibits long range entanglement. This leads to a large number of non-zero Schmidt coefficients ---and therefore large block entropy.
In contrast, for large amplitudes of the field, the second sum of Eq.~(\ref{eq:hamiltonian}) dominates the behavior of the system: the ground state is a product state with only one non-zero Schmidt coefficient
(that must be equal to one because of normalization), which gives $S(p)=0$.

\section{Numerical Results}
\subsection{Methods and Materials}
To obtain the ground state of finite chains for arbitrary configurations of disorder
we employ the Time Evolving Block Decimation algorithm with an imaginary time 
evolution \cite{Vidal2003,Vidal2004,Vidal2007}. 
The algorithm is based on calculating Schmidt decompositions at all links of the spin chain, 
which leads to describing the quantum state through a product of matrices. 
The rank of these matrices reflects the number of Schmidt coefficients that are retained 
for the simulations. Therefore, slightly entangled systems (in terms of the number of 
non-vanishing Schmidt coefficients) are described accurately by small matrices, 
which leads to a large computational speedup. 
Strongly entangled systems, in contrast, 
require very large matrices to be described accurately.
Excessive truncation of the matrices induces a breakdown of the algorithm, 
although in general one can monitor the accuracy before this happens --- 
for example, by measuring the value of the smallest retained Schmidt coefficients. 
We also perform additional tests to ensure that the numerical solution does not depend on the maximum number of Schmidt coefficients.

The TEBD algorithm used for our simulations is particularly efficient for one-dimensional 
systems with on-site and nearest-neighbors interactions only, as is the case for the $XY$ 
system in presence of the external random field. We implemented the algorithm for finite 
systems with open boundary conditions, and for infinite systems with a 
periodic Hamiltonian by imposing the periodicity of the solution. 
Due to the numerical complexity of the algorithms, we used a 
wide range of resources from desktop computers and local clusters to the 
Zaragoza supercomputer with up to 50 parallel processors.

The disorder-induced order effect is symmetric with respect to the orthogonal direction of 
the disordered field. Because of this, the original proposal used convenient 
boundary conditions in order to lift this symmetry \cite{Wehr2006}. For our 
simulations, usually it turned out to be enough to impose non-symmetric initial conditions 
for computing the imaginary time evolution towards the ground-state.

\subsection{Staggered field}
We begin by reviewing the case of an staggered magnetic field,  $h^i = (-1)^i h_0$.
Although it is not random, its non-uniformity will help us gain
a good intuition for the random case.  

\begin{figure}
\centering
\includegraphics*[width=9.6cm]{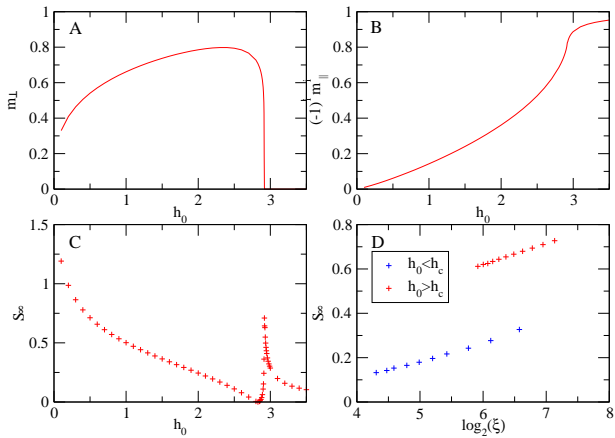}
\vskip-0.4cm
\caption{Panel A shows the ground state magnetization in the direction orthogonal to the staggered field depending on the strength of the staggered field. In B, we see the staggered magnetization in the direction parallel to the external field.
Panel C represents the entropy of entanglement $S_\infty$ for a semi-infinite block. Results in panel D show the entropy of entanglement $S_\infty$ as a function of the logarithm of the correlation length near the critical point $h_c \simeq 2.915$.}
 \label{one}
 \end{figure}

We observe two distinct regimes as a function of the magnetic field intensity $h_0$
(see Fig.~\ref{one}A).
For small fields, a finite spontaneous magnetization arises in the direction orthogonal 
to the field.
On the other extreme, for large $h_0$, the magnetization in this direction is zero.
Interestingly, we observe that the transition between the two regimes is sharp,
indicating the presence of a second order quantum phase transition.
Our numerical estimate of the critical point is $h_c=2.915 \pm 0.001$,
which is in agreement with previous studies \cite{Kurmann}.
 As the field intensity approaches $h_c$ from below, the spontaneous magnetization decays according to a power law,
$m_\perp(h_0) \sim (1-h_0/h_c)^{\beta}$. Our numerical analysis gives $\beta=0.125 \pm 0.002$.
In Fig.~\ref{one}B, we see evidence that at the critical field intensity of the staggered magnetization along the direction of the field does, indeed, show a singularity in the first derivative. 

In Fig.~\ref{one}C
we show the block entropy $S_\infty$ for a semi-infinite block as a function of intensity of the staggered field. 
Near the critical point, the entropy of a semi-infinite block diverges as \cite{Cardy}:
\begin{equation}
S_\infty = \frac{1}{2} \frac{c}{3} \textrm{log}_2(\xi) + a.
\label{entropy_infinite_block}
\end{equation}
where $c$ is the central charge of the underlying conformal field theory. 
The factor $\frac{1}{2}$ in (\ref{entropy_infinite_block}) appears 
because we measure entropy between two semi-infinite parts of the chain with only one boundary between them. 
Through a best fit to the data shown in Fig.~\ref{one}D, we obtain a value of $c = 0.53 \pm 0.05$ for the central charge.
 
For small values of the field, the entropy diverges as we approach the isotropic $XY$ critical point. For larger values of the field intensity, entropy
decays to zero, which is expected as the ground state becomes a product state.
As a curiosity, for field intensities smaller than the critical, there is a 
special value $h_0 = 2\sqrt{2}$ of the field for which the block entropy is exactly zero, 
and the ground state is thus a N\'{e}el product state \cite{Kurmann}.

\subsection{Oscillating fields}

Next, we focus on the case of a smooth periodic field such that at site $i$ the field is 
$h^i_\parallel=h \sin ( k i )$, where $k\ll1$ is the wave number of the 
periodic field. The system exhibits spontaneous perpendicular magnetization for small, 
non-zero values of $h$, whereas for large intensities $h$ the parallel magnetization follows 
the oscillating field.

\begin{figure} 
\centering
\includegraphics*[width=8.6cm]{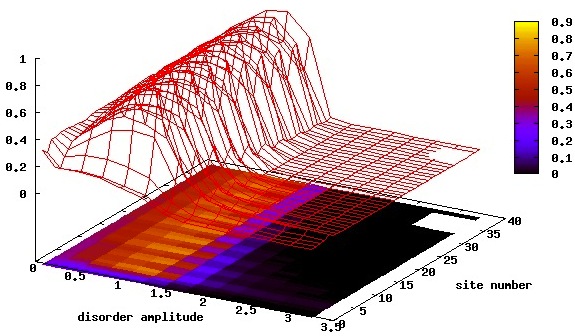}
\vskip-0.4cm \caption{Orthogonal ground state magnetization in presence of a regularly oscillating field. For $k=\frac{2 \pi}{8}$, for example, we see the appearance of a regime with orthogonal magnetization for disorder amplitudes of around 1.0 and disappearance of orthogonal magnetization around disorder amplitude of about 1.5. This confirms that uni-axially oscillating magnetic fields can induce magnetization orthogonal to the oscillating direction.}
\label{omag_osc}
\end{figure}

Figure~\ref{omag_osc} shows the orthogonal magnetization of the individual spins for different amplitudes of the external oscillating field. Similar to the staggered field (Fig~\ref{one}B), we observe two regimes of orthogonal magnetization: presence of orthogonal magnetization for small amplitudes up to a given value depending on $k$, and disappearance thereof for larger amplitudes. The sinusoidal nature of the uni-axial external magnetic field is translated to a slight variation in the strength of the orthogonal magnetization.

\begin{figure} 
\centering
\includegraphics*[width=8.6cm]{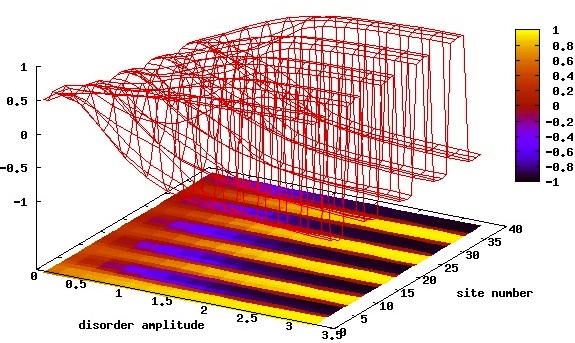}
\vskip-0.4cm \caption{Parallel magnetization in the presence of a regularly oscillating field. The plot shows that the spin chain magnetizes according to the external magnetic field if this uni-axially oscillating field has a large enough amplitude. This confirms that the physical intuition trivially works for strong magnetic fields.}
\label{cmag_osc}
\end{figure}

In Fig.~\ref{cmag_osc}, we see that the magnetization of the spins is dominated by external field when the latter has large amplitudes,  confirming physical intuition. The amplitude region of the presumed phase transition shows a strictly monotone increase in the tendency of individual spins to align with the external magnetic field.

\begin{figure} 
\centering
\includegraphics*[width=8.6cm]{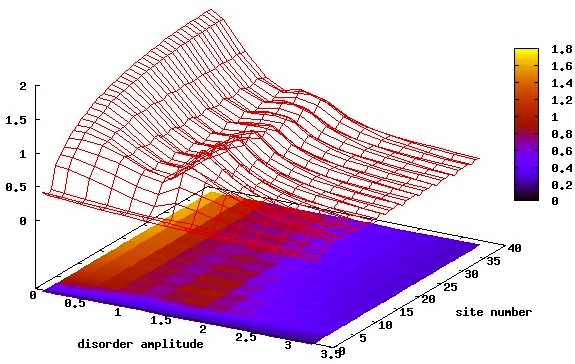}
\vskip-0.4cm \caption{Ground state block entropy of a partition in the presence of a regularly oscillating field. This configuration shows a minimum for amplitudes around 1.0 corresponding to orthogonal magnetization and a maximum for amplitudes of around 1.5 corresponding to the abrupt disappearance of the orthogonal magnetization, further indicating the presence of a quantum phase transition. This plot also indicates that the boundary effects become negligible beyond 3-5 sites from the edge of the spin chain.}
\label{cent_osc}
\end{figure}

We show the block entropy as a function of the site and disorder amplitude in Fig.~\ref{cent_osc}.
As expected, at very low fields --- near the isotropic $XY$ critical point --- the 
entropy grows slowly with system size, while it stabilizes rapidly for larger $h$. Moreover, the saturation 
value decreases with field intensity. The entropy has a local maximum at a field value that 
coincides with an abrupt decrease in perpendicular magnetization.

\begin{figure}[ht] 
\includegraphics[width=\columnwidth,clip]{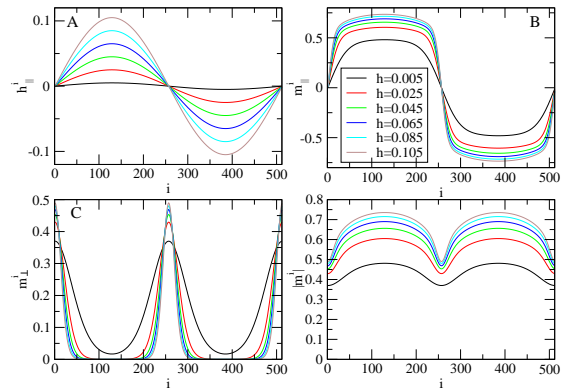}
\caption{Results for infinite system simulated with $N=512$ sites and periodic boundary conditions. Panel A shows the magnetic field $h^i_\parallel$. In B and C
respectively magnetization along the field $m_\parallel$ and spontaneous magnetization $m_\perp$. In panel D we show the total magnetization $|m^i|=\sqrt{ (m_\perp^i) ^2+(m_\parallel^i)^2 }$.}
\label{sinfMx}
\end{figure}

Figure~\ref{sinfMx} summarizes our studies of the oscillating fields with different amplitudes. Comparing the magnetization in Fig.~\ref{sinfMx}A to the transverse magnetization shown in Fig.~\ref{sinfMx}C, we see that spontaneous symmetry breaking appears near the zeros of $h^i_\parallel$. For strong enough intensities $h$, this leads to the creation of a set of ``islands'' of perpendicularly magnetized spins in a sea of transverse magnetization. When $h$ is weak enough, however, the island size $R$ becomes greater than the distance between the islands $\pi/k$, the isolated islands merge, and there is non-zero $m_\perp$ of definite sign everywhere (Fig.~\ref{sinfMx}C). The periodic transition from one phase to the other can be understood 
in terms of quantum phase transitions in space \cite{space}: 
away from the critical points, the system follows the local value of the field
adiabatically (Fig.~\ref{sinfMx}B) and remains on the corresponding phase.
In our system, this corresponds to the regions where the field is 
very large and therefore the local magnetization is in the symmetric phase $m^i_\perp=0$ (see \cite{McCoy}). However, when the amplitude of the field $h^i_\parallel$ approaches its critical value, 
the local correlation length of the system,
$\xi\simeq |h^i_\parallel |^{-\nu}$,
can become much larger than the rate of change of the field, 
$\ell \simeq \left|h^i_\parallel/\frac{d h^i_\parallel}{di}\right|$. 
In this regime  the system cannot heal fast enough (compared to the change in the field), 
and it begins to transition from one phase to the other, forming an island 
of broken symmetry with a random sign of $m_\perp$. 
We can estimate the size $R$ of an island by linearizing the magnetic field near its 
zero at $i_0$: $|h^i_\parallel|\approx |h k \: (i-i_0)|$.
At the boundary of the island we have the condition 
$\xi \simeq \ell$,
which writes as
$| h k \: R |^{-\nu}\simeq| R |$. Thus, the size of an isolated island of perpendicular magnetization
results
\be
R \simeq (h k)^{-\frac{\nu}{\nu+1}}.
\label{deltai}
\ee 
This line of reasoning, based on the Kibble-\.{Z}urek mechanism, allows us to 
also estimate how the amplitude of perpendicular magnetization in 
the islands depends on $h$ and $k$. 
In a first approximation, the total magnetization is constant during the transition 
in space (Fig.~\ref{sinfMx}D). 
When the local density approximation starts to break, i.e. at $i_0 \pm R$,
the system goes from an adiabatic to an impulse region, and the order parameter at this point must
``freeze''. Therefore, we can estimate the amplitude of perpendicular magnetization in the island 
with the value of the parallel magnetization at the freezing point, 
$m_\perp^{ i_0} \sim |m_\parallel^{i_0\pm R} |$. Using that near the critical point 
$|m^i_\parallel| \sim |h^i_\parallel|^{1/\delta}$ \cite{McCoy}, we obtain
\begin{equation}
m_\perp^{ i_0}~\sim~\left(h k R\right)^{1/\delta}~=~(h k)^{\frac{1}{\delta(\nu+1)}}. 
\end{equation}  

We simulated an infinite system with $k = 2 \pi / N$ and N=512. We compare with the critical
exponents of a pure $XY$ chain, $1/\delta\simeq 0.14$ and $\nu \simeq 0.57$ \cite{McCoy}.
The results, shown in 
Fig.~\ref{sinfMx}, give $R \simeq h^{-0.369}$, with the exponent
close to the predicted value $\frac{\nu}{\nu+1} \simeq 0.363$. 
For the amplitude of the magnetization  of the islands we 
obtain $m_\perp^{i_0} \sim h^{0.092}$, again in good agreement with our prediction 
$\frac{1}{\delta(\nu+1)}\simeq 0.089$.

\subsection{Randomly oscillating fields}
\begin{figure} 
\centering
\includegraphics*[width=8.6cm]{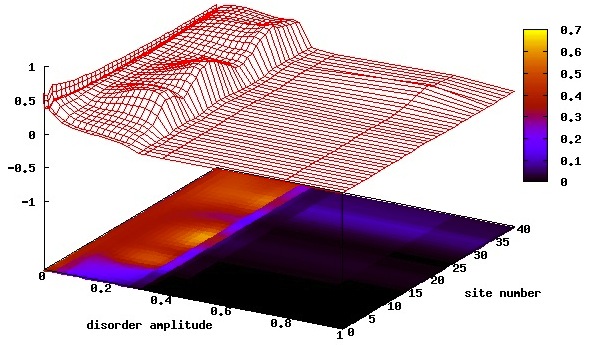}
\vskip-0.4cm \caption{Orthogonal magnetization of a particular realization of the random external field. We see that magnetization in the orthogonal direction of the random field is non-zero up to a threshold. The sudden drop to zero at a certain amplitude (0.3 in this realization) indicates the phase-transition between the disorder-induced order state and the state primarily following the external magnetic field.}
\label{omag_ds}
\end{figure}

\begin{figure} 
\centering
\includegraphics*[width=8.6cm]{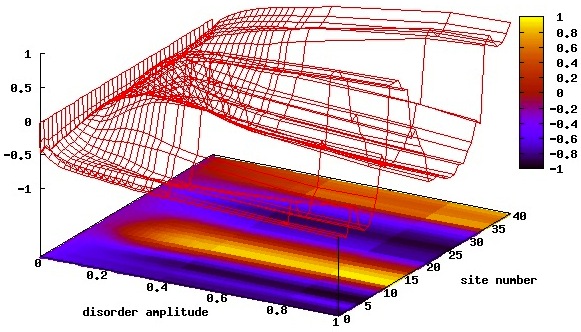}
\vskip-0.4cm \caption{Parallel magnetization for a given random field. Note that no abrupt change occurs in the amplitude region of the phase-transition (here 0.3). At unit amplitude, the parallel magnetization is no longer exactly following the quick variations of the randomly oscillating external field.}
\label{cmag_ds}
\end{figure}

We study randomly oscillating fields produced from a normal distribution of mean zero and varying standard deviation. A large standard deviation is equivalent to a large amplitude of an oscillating field. Producing pseudo-random numbers following a normal distribution is similar to the possible experimental realization of a disordered field using laser speckles (which has already been demonstrated).
Figure~\ref{omag_ds} shows the formation of islands of magnetization, coinciding with positive, negative, or alternating regions of the parallel magnetization. Since the external random field oscillates rapidly, the parallel magnetization, shown Fig.~\ref{cmag_ds}, cannot no longer follow the field exactly, even at unit amplitude.

\begin{figure} 
\centering
\includegraphics*[width=8.6cm]{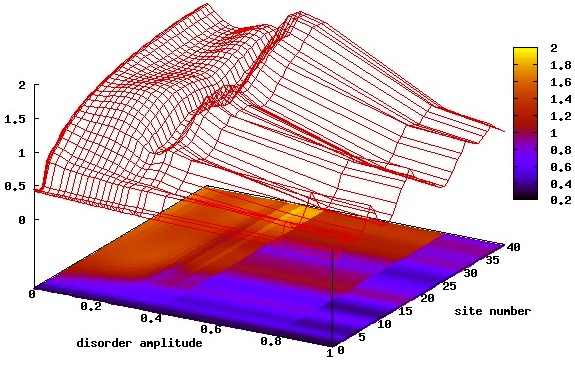}
\vskip-0.4cm \caption{Block-entropy for a particular realization of the pseudorandom external field. Note the marked minimum shortly before, and the maximum at the disorder-amplitude for which the orthogonal magnetization disappears (0.3 in this realization of the disorder). This apparent discontinuity in the block entropy also indicates the presence of a quantum phase transition between an orthogonally magnetized and a not magnetized state. }
\label{cent_ds}
\end{figure}

In close analogy to the staggered and sinusoidally oscillating field, Fig.~\ref{cent_ds} shows a maximum of the block entropy at the disorder amplitude corresponding to the disappearance of the orthogonal magnetization. In fact, the bulk of the material no longer shows a monotone decrease of the block entropy for increasing amplitudes up to the disappearance of the orthogonal magnetization. There now appears a more complex structure dumping into a marked minimum right before the maximum at which the amplitude of the orthogonal magnetization vanishes. The contrast between these two final extremal points resembles the discontinuity observed in the staggered field and appears to support the claim of a phase-transition at the corresponding disorder amplitudes.

The fact that even for uncorrelated magnetic fields there appears to be (in most cases), a region of orthogonal magnetization illustrates the robustness of the effect. Our studies show that the disorder amplitude for which the orthogonal magnetization disappears now strongly depends on the individual uni-axial random field configuration. 

\begin{figure} 
\centering
\includegraphics*[width=8.6cm]{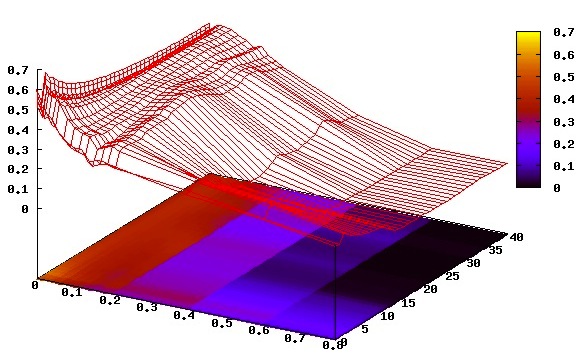}
\vskip-0.4cm \caption{Average orthogonal magnetization in the bulk of the spin chain. Our results show the existence of a range of disorder amplitudes, for which disorder-induced order occurs. The maximum average orthogonal magnetization is obtained for disorder strengths of approximately 0.3; the average is taken over 10 realizations of pseudorandom external fields.}
\label{avg_orthmag_ds}
\end{figure}

Figure~\ref{avg_orthmag_ds} shows the mean value over 10 disordered realizations
of the perpendicular magnetization. The graph shows a clear average presence of orthogonal magnetization for small amplitudes. The mean orthogonal magnetization appears to be strongest for a disorder amplitude of approximately $30\%$ of the $XY$ spin-spin correlation.

As expected, the randomly oscillating field presents a number of islands of different sizes because there are regions of predominantly positive, negative or oscillating random values. When averaging over several realizations, we still see a clear overall induced constant order for small amplitudes of the external magnetic uni-axial random field.

\section{Experimental realization}

The proposal below combines Raman coupling to realize the spin dependent lattice with radio-frequency transitions to induce the desired structure of the hopping matrices. The main idea stems from L.~Mazza {\it et al.} \cite{mazza}, and consists of creating a lattice that traps certain bosonic alkali atoms in all states from the lower hyperfine manifold,   and atoms in a  certain state of the upper hyperfine manifold in between the sites of the square lattice (the original goal of this approach was to design non-Abelian gauge fields in 3D lattice, details will be published elsewhere). 
	
The main ingredient is to use the \textit{magic wavelength}, experimentally realized in systems with $^{87}$Rb. This is in  principle feasible with all the alkaline atoms, since they share the same fine structure. We propose to use the wavelength $\bar{\lambda}$ such that the contribution for trapping the state $\ket{S_{1/2}; m_S = +1/2}$ coming from the states $\ket{P_{3/2}; m_S = -1/2}$ and $\ket{P_{1/2}; m_S = -1/2}$ exactly cancel. In this way, the state $\ket{S_{1/2}; m_S =1/2}$ feels the potential coming from the $\sigma_+$ polarized light whereas the state $\ket{S_{1/2}; m_S=-1/2}$ that coming from the $\sigma_-$ polarized light (the quantization axis of $m_S$ coincides with the propagation direction of the circularly polarized light).
One should avoid  working with light atoms such as $^7$Li due to the small fine splitting. We can, however, take heavier atoms (like $^{39}$K,  $^{41}$K or even better  $^{85}$Rb or $^{87}$Rb) and eventually pump the atoms to the extremal Zeeman levels, that will then serve as spinless atoms.  This approach  will limit the lifetime to be $\le 1$s, but should suffice to observe at least some of the physics of DIO. For experimental realizations along these lines, see for example~\cite{BAPS}.

We start by confining the atoms in 1D, say along the $X$ axis. We take  two laser pulses propagating in the $X$ direction with circularly polarized light with respect to the  $X$ direction.  
Denoting by $I$ the nuclear spin, the general result of Ref. \cite{mazza} is that for appropriately designed laser fields  all $F=I-1/2$ states are trapped in the 1D lattice sites, with $X$ now the natural quantization axis. The optical potential for the $F=I+1/2$ manifold has  minima in the same lattice sites, but, interestingly, develops also minima for the states $|F=I+1/2;F_x=I+1/2\rangle$  in the middle of the links in the $X$ directions. These states, on the one hand,  have good overlaps with the  $F=I-1/2$ states in the basic 1D lattice sites, and obviously can serve as intermediate states for the radio-frequency transitions between the $F=I-1/2$ atoms. On the other hand, the trapping potential for the states in the $F=I+1/2$ manifold  can be quite weak, so that tunneling effectively dominates over interactions. The trapping potential for $F=I-1/2$ atoms can be strong enough to put them in the Mott insulating regime. 

In the case of Rubidium $I=3/2$, one should prepare a large condensate in the $F=2$ manifold, and then pump some atoms to the $F=1,F_x=-1$ states. In the strong repulsion regime (hard bosons regime), the Hamiltonian for $F=1$ atoms reduces to that of the $XY$ model. The uni-axial random field in the $XY$ plane can be easily realized using Raman (optical or RF) transitions with fixed phases and random strengths, as in the proposal of Ref. \cite{Niederberger2008}. 

Interestingly, the same scheme can be generalized to 2D in a square, and even 3D in a simple cubic lattice.  In 2D for instance, we start by confining the atoms in 2D, say in the $XY$plane. We take  two laser pulses propagating in the $X$ direction with circularly polarized light with respect to the  $X$ direction.  Similarly, we apply two laser pulses in the $Y$ direction with the circular polarizations corresponding to propagation axis $Y$. 
Again, the general result is that for appropriately designed laser fields  all $F=I-1/2$ states are trapped in the square lattice sites. The natural quantization axis for them is now a 45 degrees axis between $X$ and $Y$. 
The optical potential for the $F=I+1/2$ manifold has  minima in the same lattice sites, but, interestingly, develops also minima for the states $|F=I+1/2;F_x=I+1/2\rangle$ and $|F=I+1/2;F_y=I+1/2\rangle$ in the middle of the links in the $X$ and $Y$ directions, respectively. These states  have good overlaps with the  $F=I-1/2$ states in the basic square lattice sites, and obviously can serve as intermediate states for the radio-frequency transitions between the $F=I-1/2$ atoms. The remaining ingredient of the proposal  are  the same as in 1D.

\section{Conclusions}

After the effect of disorder-induced order has been shown in classical systems, we have presented numerical evidence that this effect also exists in quantum systems such as quantum $XY$ chains. The effect consists in the appearance of magnetization in the direction orthogonal to a spatially disordered external magnetic field for small amplitudes of this field. 
The key ingredients for justifying this result are values of various components of magnetization (which are directly measurable in the experimental scheme discussed above), and non-monotone block entropy. The latter cannot be measured directly, but its properties can be inferred from the measurements of density-density correlations using Bragg spectroscopy, noise interferometry, and/or spin polarization spectroscopy.  Finally, let us mention that recently an analogue of DIO in the time domain (i.e. with time dependent perturbations) has been proposed and termed {\it rocking} \cite{Stali}.

A.N. acknowledges the many fruitful discussions with Alex Cojuhovschi during the development of the numerical code. J.W. thanks L.~Torner and ICFO for their hospitality and support during the summers of 2008 and 2009. The authors thankfully acknowledge the computer resources, technical expertise and assistance provided by the Barcelona Supercomputing Center - Centro Nacional de Supercomputaci\'{o}n under grants FI-2008-3-0029 and FI-2009-1-0019 for the Zaragoza computers. Also, we acknowledge financial support from the Spanish MINCIN project FIS2008-00784 (TOQATA), Consolider Ingenio 2010 QOIT, Polish Government  research projects N202 175935 (J.D.) and N202 174335 (M.M.R), EU STREP project NAMEQUAM, ESF QUDEDIS grant no. 1017 (M.M.R.) and exchange grant no. 1759 (A.N.), ERC Advanced Grant QUAGATUA, Caixa Manresa Chair, U.S.~Department of Energy through the LANL/LDRD Program (M.M.R.),  and from the Humboldt Foundation.

\end{document}